\documentstyle[12pt,cites,epsf,rotate,amssymb,amsfonts]{article}
\begin{document}

\title{\bf Dependence of the Excitation Energies of Boron in
Diamond on Isotopic Mass}
\author{Manuel Cardona\thanks{Email: cardona@kmr.mpi-stuttgart.mpg.de, Fax:
+49-711-689-1712}
\\[18pt]
Max-Planck-Institut f\"ur Festk\"orperforschung, \\
Heisenbergstr.\ 1, 70569 Stuttgart, Germany, European Union}
\date{Received September 19, 2001 by J. Kuhl\\Accepted Sept. 25, 2001 for Solid State Commun.}
\maketitle

\noindent {\bf Abstract}\\[0.5cm] \noindent Kim {\it et al.} have
reported a dependence of the infrared  excitation energies of
boron acceptors in diamond on the isotopic mass of the carbon
atoms. We show that this change can be quantitatively interpreted
as induced by a change in the hole effective mass.\\[0.5cm]

\section{Introduction}
Kim {\it et al.} have investigated the dependence of the infrared
excitation spectrum of substitutional boron acceptors in diamond
on the isotopic mass of diamond ($M_i\simeq$12 for a natural
diamond, $M_i\simeq 13$ for a synthetic diamond grown out of
$^{13}$C) \cite{kim}. Six absorption peaks with average energy
equal to 351 meV were observed (see peaks 12a,13,14,15,20 and 22
in Fig.~3 of \cite{kim}; peak 11 is very weak and will not be
included in our considerations). They discovered an increase of
1.26 meV in the average energy of these peaks when replacing
$^{12}$C by $^{13}$C. They examined various possible origins for
this change (e.g., change of the dielectric constant with $M_i$
\cite{ruf}, changes of a self-energy related to electron-phonon
interaction \cite{coll}) but were not able to give any convincing
explanation for the origin of that dependence on $M_i$. In this
work, we attribute the change in excitation energy to a change in
the binding energy of the acceptor related to a change in the hole
effective mass. The latter results from the renormalization of the
$E'_0$ gap of diamond by electron-phonon interaction \cite{zoll}.
The $E'_0$ gap determines, via {\bf k.p} perturbation theory, the
average hole mass which defines the acceptor binding energy.

\section{Theory}
The average excitation energy under consideration (351 meV)
 is quite close to the ionization energy of the boron
acceptor (370 meV) \cite{mass}. We shall therefore assume that the
measured relative change in average excitation energy, $(\Delta
E_{ex}/E_{ex})= +3.6\times 10^{-3}~~[\Delta E_{ex} =
E_{ex}(^{13}\rm {C}) - E_{ex}(^{12}\rm{C})]$, is equal to the
corresponding change in binding energy. For nondegenerate band
extrema the binding energy of a hydrogenic impurity is:

\begin{equation} 
E_B ~=~ 13.6 ~~\frac{m^*}{\epsilon_0^2}~~{\rm (eV)}~ ,
\end{equation}

\noindent where $m^*$ is the effective mass (in units of the free
electron mass) and $\epsilon_0\simeq 5.7$ for diamond. For
electrons in conventional semiconductors, $m^*$ is, to a good
approximation, proportional to an interband gap \cite{yu}. For
holes, however, the relevant bands are degenerate at ${\bf k} =0$
and the acceptor binding energy is determined by the so-called
Luttinger parameters, labeled $\gamma_1, \gamma_2$ and $\gamma_3$
(see \cite{yu}, p. 174). These parameters are not well known for
diamond. We use here the theoretical values $\gamma_1 = 2.54,
\gamma_2\simeq 0, \gamma_3 = 0.63$ reported in \cite{willa}. The
dominance of $\gamma_1$ enables us to use for $E_B$ the ``one
spherical band'' approximation (\cite{yu}, p. 176) and to write:
\begin{equation} 
E_B~ = ~~\frac{13.6}{\gamma_1~ \epsilon_0^2} ~~\rm (eV) .
\end{equation}

It is easy to see, using the data of \cite{ruf}, that the
dependence of $\epsilon_0$ on $M$ is an order of magnitude too
small to account for the observed dependence of $E_{ex}(\simeq
E_B)$. Let us therefore consider the dependence of $\gamma_1$ on
$M$. For this purpose, we use the expression \cite{willa}:

\begin{equation} 
\gamma_1 =~ -\frac{1}{3}~~(2F~+~4G~ +~ 4M) -1
\end{equation}

\noindent which for the definition and values of $F, G$, and $M$
given in \cite{willa} can be written as:

\begin{equation} 
\gamma_1\simeq~ -\frac{4M}{3} ~~=~~\frac{Q^2}{E'_0}~ ,
\end{equation}

\noindent where $E'_0$ represents the lowest direct gap of diamond
and $Q$ the matrix element of linear momentum connecting the
valence and conduction states which define $E'_0$. This matrix
element is closely related to $2\pi/a_0$, where $a_0$ is the
lattice parameter which also depends only weakly on $M$
\cite{pav}. We thus assume, as is often done in the spirit of {\bf
k.p} theory, that the dependence of $\gamma_1$ on $M$ results from
the corresponding dependence of the $E'_0$ gap.

\section{Results and Discussion}
Combining Eqs.~(2) and (4) we obtain:

\begin{equation} 
\frac{\Delta E_{ex}}{E_{ex}} ~\simeq~ \frac{\Delta
E_B}{E_B}~=~\frac{\Delta E'_0}{E'_0}
\end{equation}

Fortunately, a calculated value of the zero-point renormalization
of the  $E'_0$ gap of diamond by the electron-phonon interaction
has been reported in \cite{zoll}. No experimental data are
available for the  $E'_0$ gap of diamond, but data for its
indirect gap and various gaps of other semiconductors support the
theoretical predictions \cite{card}. From the zero-point
renormalization ($\simeq -665$ meV for diamond) and its expected
proportionality to $M^{-1/2}$ \cite{zoll,card}, we find for
$\Delta E'_0 = E'_0 (^{13}{\rm C}) - E'_0 (^{12}\rm C)$:

\begin{equation} 
\Delta  E'_0 ~=~ \frac{1}{2} \times \frac{1}{12.5}\times 665~\rm
meV ~=~ 27~\rm meV .
\end{equation}

\vspace{0.5cm} \noindent Replacing this value of $\Delta E'_0$ and
$E'_0 = 7.2$ eV in Eq.~(4) we obtain:

\vspace{0.5cm}

\begin{equation} 
\frac{\Delta E_{ex}}{E_x} ~\simeq~ \frac{27\times10^{-3}}{7.2}~=~
3.7\times 10^{-3} ~ ,
\end{equation}

\noindent in excellent agreement with the measurements reported in
\cite{kim}($3.6\times 10^{-3})$.

\section{Conclusions}

We have identified the mechanism responsible for the isotope
effect on the excitation and binding energy  of acceptors in
diamond as being due to the renormalization of the $E'_0$ gap by
the electron-phonon interaction, which is particularly large in
the case of diamond.

\section{Acknowledgements}
I would like to thank M.L.W. Thewalt for bringing this effect to
my attention.

\end{document}